\documentclass[journal,twocolumn]{IEEEtran}

\usepackage{amsmath}
\usepackage{graphicx}
\usepackage{cite}
\usepackage{multirow}
\usepackage{booktabs}
\usepackage{amssymb}
\usepackage{enumitem}
\usepackage{url}
\usepackage{hyperref}
\usepackage{orcidlink}
\usepackage{xcolor}
\begin{document}

\title{DeRA-MOS: Optimizing Text-to-Music Evaluation via Decoupled Listwise Ranking and Modality Alignment}

\author{
Chien-Chun~Wang\,\orcidlink{0009-0000-1392-0058},~\IEEEmembership{Member,~IEEE,}
Hung-Shin~Lee\,\orcidlink{0000-0001-7044-9434},
Hsin-Min~Wang\,\orcidlink{0000-0003-3599-5071},~\IEEEmembership{Senior Member,~IEEE,}
and~Berlin~Chen\,\orcidlink{0000-0003-0693-8932},~\IEEEmembership{Member,~IEEE}

\thanks{Chien-Chun Wang is with E.SUN Financial Holding Co., Ltd., Taipei 10546, Taiwan (e-mail: jethrowang0531@gmail.com).}
\thanks{Hung-Shin Lee is with United Link Co., Ltd., Taipei 11493, Taiwan (Corresponding author; e-mail: hungshinlee@gmail.com).}
\thanks{Hsin-Min Wang is with the Institute of Information Science, Academia Sinica, Taipei 11529, Taiwan (e-mail: whm@iis.sinica.edu.tw).}
\thanks{Berlin Chen is with the Department of Computer Science and Information Engineering, National Taiwan Normal University, Taipei 11677, Taiwan (Corresponding author; e-mail: berlin@ntnu.edu.tw).}
}

\maketitle

\begin{abstract}

Evaluating text-to-music (TTM) systems remains expensive because music impression (MI) and text alignment (TA) scores rely on human mean opinion scores (MOS). Most automatic MOS estimators are trained with point-wise regression or distributional classification. These objectives do not directly optimize rank-based metrics and provide weak geometric constraints for cross-modal coherence. To address these gaps, we propose DeRA-MOS, a decoupled optimization framework for TTM evaluation. For MI, we introduce a batch-aware listwise ranking loss that models relative order within each mini-batch and better aligns with evaluation based on Spearman's rank correlation coefficient (SRCC). For TA, we introduce a score-anchored modality alignment loss that maps human scores to target audio-text similarity and regularizes the latent space before fusion. By effectively mitigating the point-wise training mismatch and modality drift, experiments on MusicEval demonstrate that our decoupled framework yields substantial improvements in both MI and TA ranking metrics, establishing a robust paradigm for large-scale TTM evaluation.

\end{abstract}

\begin{IEEEkeywords}

Text-to-music evaluation, mean opinion score, listwise ranking, cross-modal alignment, metric learning.

\end{IEEEkeywords}

\IEEEpeerreviewmaketitle

\section{Introduction}

\IEEEPARstart{T}{he} proliferation of text-to-music (TTM) generative models \cite{liu2023,agostinelli2023,copet2023,huang2023} enables high-fidelity synthesis from natural language, yet evaluation lags behind.
Quality assessment remains subjective, relying heavily on human mean opinion scores (MOS) for music impression (MI) and text alignment (TA) \cite{liu2025}.
Given costly annotations, accurate automatic MOS estimation is essential for scalable TTM development \cite{mittag2021,saeki2022}.

To automate assessment, previous efforts framed MOS prediction as regression.
Traditional approaches use point-wise objectives (e.g., L1 or MSE) to predict single scalars, standard in speech synthesis (e.g., MOSNet \cite{lo2019}, MBNet \cite{leng2021}, UTMOS \cite{saeki2022}).
Recognizing human ratings as discrete and ordinal, recent methods shifted toward distributional \cite{wong2023} and listener-dependent modeling \cite{huang2022}.
For instance, DeePMOS-$\mathcal{B}$ \cite{liang2024} uses Beta distributions capturing annotator variance, while CORAL \cite{cao2020} predicts cumulative probabilities.
Recently, DORA-MOS \cite{ritter-gutierrez2025} proposed Gaussian label softening over discrete bins to implicitly encode MOS ordinality.
Similarly, objective TA evaluation frequently uses reference-free metrics like Fréchet audio distance (FAD) \cite{kilgour2019,gui2024}.
Contemporary models often use pre-trained joint embeddings (e.g., CLAP \cite{wu2023}, MuQ \cite{zhu2025}) and cross-attention to implicitly predict coherence.

Despite these improvements, a mismatch persists between training objectives and evaluation metrics.
First, metrics like Spearman's rank correlation coefficient (SRCC) evaluate relative order, not absolute scores.
Even with ordinal smoothing \cite{geng2016,diaz2019}, point-wise training treats samples independently, ignoring batch-level ranking structure.
Second, cross-attention-based TA estimation lacks explicit geometric constraints on unimodal embeddings.
Unanchored to human judgment, these representations can drift, degrading semantic alignment.

To address these challenges, we propose \textbf{DeRA-MOS} (\textbf{De}coupled \textbf{R}anking and \textbf{A}lignment)\footnote{Our code is available at: \url{https://github.com/JethroWangSir/DeRA-MOS}.}, a framework decoupling training objectives into two specialized losses while keeping the backbone unchanged. 
Though inspired by established metric learning, our novelty lies in their targeted, decoupled combination to address TTM evaluation bottlenecks.
For MI, the Batch-Aware Listwise Ranking (BALR) loss treats each mini-batch as a query list to optimize global ranking via learning-to-rank principles \cite{cao2007,xia2008}.
For TA, the Score-Anchored Modality Alignment (SAMA) loss aligns pre-fusion cosine similarity of L2-normalized embeddings with human TA MOS \cite{guzhov2022}.
This formulation directly targets ranking metrics while stabilizing cross-modal geometry.

In summary, our main contributions are:
\begin{enumerate}[noitemsep,leftmargin=*]
\item \textbf{Batch-Aware Listwise Ranking:}
We replace point-wise MI supervision with batch-aware listwise ranking, directly aligning with SRCC evaluation.
\item \textbf{Score-Anchored Modality Alignment:}
We anchor latent audio-text similarity to TA MOS, mitigating representation drift and enhancing semantic coherence.
\item \textbf{Robust Performance with DeRA-MOS:}
On MusicEval, the unified DeRA-MOS framework consistently outperforms a reproduced homogeneous point-wise baseline in terms of SRCC for both MI and TA tasks.
\end{enumerate}

\section{Proposed Methodology}

\begin{figure}[t]
\centering
\includegraphics[width=1.0\linewidth]{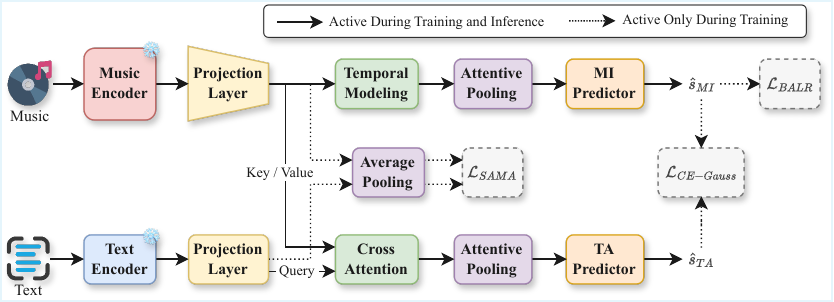}
\vspace{-22pt} 
\caption{
\textbf{Overview of the proposed DeRA-MOS framework.}
The architecture explicitly decouples the TTM evaluation process using task-specific objectives (dotted paths denote training-only operations).
To overcome the limitations of homogeneous point-wise learning ($\mathcal{L}_{CE-Gauss}$), we introduce $\mathcal{L}_{BALR}$ to enforce batch-aware global ranking for the MI branch.
Crucially, for the TA branch, $\mathcal{L}_{SAMA}$ is applied \emph{before} cross-attention fusion to geometrically anchor the unimodal representations, preventing latent drift and ensuring robust cross-modal coherence.
}
\label{fig:main}
\vspace{-15pt}
\end{figure}

While DORA-MOS effectively models TTM evaluation via a dual-branch structure, its optimization objective is sub-optimal.
Specifically, it uses a point-wise Gaussian-softened classification loss.
Though capturing MOS's ordinal nature, it treats samples independently, ignoring batch-level ranking dynamics.
Furthermore, for TA, it learns modality coherence implicitly via cross-attention, lacking a direct geometric constraint.
We introduce two complementary objectives that explicitly enforce global ranking and cross-modal geometric consistency, as illustrated in Fig. \ref{fig:main}.

\subsection{Batch-Aware Listwise Ranking for Music Impression}

The evaluation of music impression is inherently a relative ranking task.
A model predicting shifted absolute values can still achieve a perfect SRCC if the rank order is preserved.
Point-wise objectives fail to optimize this metric directly.
To reduce the mismatch between training and SRCC-oriented evaluation, we propose a batch-aware listwise ranking loss.
Inspired by foundational learning-to-rank paradigms such as ListNet \cite{cao2007,xia2008} and LambdaRank \cite{burges2006}, we conceptualize the entire mini-batch as a single query list.
We transform both ground-truth and predicted MI scores into batch-wise probability distributions using temperature-scaled softmax.
Let $\mathbf{s}_{MI} = [s_{MI}^{(1)}, s_{MI}^{(2)}, \dots, s_{MI}^{(B)}]^T$ be the ground-truth MI scores for a batch of size $B$, and $\hat{\mathbf{s}}_{MI} = [\hat{s}_{MI}^{(1)}, \hat{s}_{MI}^{(2)}, \dots, \hat{s}_{MI}^{(B)}]^T$ be the corresponding model predictions.
For sample $i$, the target and predicted listwise probabilities are defined as
\vspace{-3pt}
\begin{equation}
p_i = \frac{\exp(s_{MI}^{(i)} / \tau)}{\sum_{j=1}^B \exp(s_{MI}^{(j)} / \tau)}, \quad \hat{p}_i = \frac{\exp(\hat{s}_{MI}^{(i)} / \tau)}{\sum_{j=1}^B \exp(\hat{s}_{MI}^{(j)} / \tau)},
\end{equation}
where $\tau > 0$ is the temperature.
When MOS values are densely concentrated (e.g., between 3.0 and 4.0), $\tau$ controls the sharpness of the induced distribution and therefore the sensitivity to small rank differences.
The BALR loss is defined as the cross-entropy between $\{p_i\}_{i=1}^{B}$ and $\{\hat{p}_i\}_{i=1}^{B}$:
\vspace{-3pt}
\begin{equation}
\mathcal{L}_{BALR} = - \sum_{i=1}^B p_i \log \hat{p}_i.
\end{equation}
Minimizing $\mathcal{L}_{BALR}$ encourages correct global ordering within each batch, rather than only absolute score matching, and thus provides gradients more consistent with SRCC.
Theoretically, this continuous approximation translates discrete rank metrics into differentiable objectives. Unlike pairwise methods (e.g., QAMRO \cite{wang2025b}) that evaluate $\mathcal{O}(B^2)$ isolated local pairs without full list context, BALR elegantly captures global batch dynamics while reducing complexity to $\mathcal{O}(B)$.

\subsection{Score-Anchored Modality Alignment for Text Alignment}
\label{sec:sama}

For predicting the TA score, previous methods fuse audio and text representations using cross-attention, assuming the network will implicitly discover semantic correspondence.
However, without an explicit constraint on the individual unimodal spaces before fusion, the embeddings can drift, leading to representations that memorize the training set rather than learning true modality coherence.
To regularize the latent space, we propose the score-anchored modality alignment loss.
The key idea is to align latent audio-text similarity with human TA MOS before cross-attention fusion \cite{chen2020}.
Let $\mathbf{h}_a^{(i)}$ and $\mathbf{h}_t^{(i)}$ denote the pre-fusion audio and text embeddings for sample $i$, which are obtained by applying average pooling over their respective temporal features.
First, we compute cosine similarity as
\vspace{-5pt}
\begin{equation}
\operatorname{sim}(\mathbf{h}_a^{(i)}, \mathbf{h}_t^{(i)}) = \frac{\mathbf{h}_a^{(i)} \cdot \mathbf{h}_t^{(i)}}{\|\mathbf{h}_a^{(i)}\|_2 \|\mathbf{h}_t^{(i)}\|_2}.
\end{equation}
Since cosine similarity ranges from $-1$ to $1$, we linearly transform it to the range $[0, 1]$ using $\tilde{c}^{(i)} = (\operatorname{sim}(\mathbf{h}_a^{(i)}, \mathbf{h}_t^{(i)}) + 1)/2$.
Correspondingly, we linearly map the human-annotated TA MOS $s_{TA}^{(i)}$ (bounded within $[1, 5]$) to $y_{sim}^{(i)} \in [0, 1]$:
\vspace{-2pt}
\begin{equation}
\label{eq:sim}
y_{sim}^{(i)} = \frac{s_{TA}^{(i)} - 1.0}{4.0}.
\end{equation}
The SAMA loss is then defined as
\vspace{-3pt}
\begin{equation}
\mathcal{L}_{SAMA} = \frac{1}{B} \sum_{i=1}^B \left( \tilde{c}^{(i)} - y_{sim}^{(i)} \right)^2.
\end{equation}
This parameter-free mapping empirically outperforms nonlinear alternatives.
Empirical sensitivity analysis further confirmed that adopting L2 normalization and explicitly anchoring to the $[0, 1]$ range prevented the instability and divergence observed in unnormalized mappings.
It bounds the space and mitigates drift without overfitting.

\subsection{Joint Optimization}

The overall training objective combines our proposed losses with the baseline task-specific losses.
Let $\mathcal{L}_{CE-Gauss}$ denote the sum of the baseline Gaussian-softened classification losses for both the MI and TA branches.
The final loss function is defined as
\vspace{-5pt}
\begin{equation}
\mathcal{L}_{\mathrm{Total}} = \mathcal{L}_{CE-Gauss} + \alpha \mathcal{L}_{BALR} + \beta \mathcal{L}_{SAMA},
\end{equation}
where $\alpha \ge 0$ and $\beta \ge 0$ are weighting coefficients balancing the contribution of the global ranking constraint and the geometric alignment constraint, respectively.
During training, $\mathcal{L}_{BALR}$ is applied to MI predictions, while $\mathcal{L}_{SAMA}$ is applied to pre-fusion TA embeddings.
During inference, both auxiliary losses are removed, and the prediction pipeline remains identical to the baseline.

\begin{table*}[t]
\centering
\caption{
Performance evaluation on MusicEval.
Models marked with an asterisk (*) denote official results reported by the challenge organizers or respective authors, while unmarked models indicate our locally reproduced baselines and proposed variants.
For MSE, lower is better ($\downarrow$); for all other metrics, higher is better ($\uparrow$).
Best overall results are highlighted in \textbf{bold}.
}
\vspace{-8pt}
\label{tab:main_results}
\setlength{\tabcolsep}{11.5pt}
\begin{tabular}{l|cccc|cccc}
\toprule
\multirow{2}{*}{\textbf{Model}} & \multicolumn{4}{c|}{\textbf{Music Impression}} & \multicolumn{4}{c}{\textbf{Text Alignment}} \\
\cmidrule{2-9}
 & \textbf{MSE $\downarrow$} & \textbf{LCC $\uparrow$} & \textbf{SRCC $\uparrow$} & \textbf{KTAU $\uparrow$} & \textbf{MSE $\downarrow$} & \textbf{LCC $\uparrow$} & \textbf{SRCC $\uparrow$} & \textbf{KTAU $\uparrow$} \\
\midrule
% Official Data Section
MusicEval-Baseline* \cite{liu2025} & 0.378 & 0.821 & 0.818 & 0.623 & 0.199 & 0.744 & 0.724 & 0.532 \\
DRASP* \cite{yang2025} & 0.076 & 0.949 & 0.957 & 0.858 & 0.058 & 0.897 & 0.890 & 0.726 \\
QAMRO* \cite{wang2025b} & 0.139 & 0.961 & 0.972 & 0.876 & 0.109 & 0.918 & 0.916 & 0.763 \\
DORA-MOS* \cite{ritter-gutierrez2025} & \textbf{0.017} & 0.986 & 0.988 & 0.913 & 0.033 & 0.946 & 0.944 & 0.809 \\
\midrule
% Reproduced and Proposed Models Section
DORA-MOS (Reproduced) & 0.018 & 0.985 & 0.981 & 0.890 & 0.060 & 0.956 & 0.952 & \textbf{0.835} \\
\quad + Ranking ($\mathcal{L}_{BALR}$) & 0.030 & 0.985 & 0.985 & 0.908 & 0.031 & 0.946 & 0.940 & 0.789 \\
\quad + Alignment ($\mathcal{L}_{SAMA}$) & 0.031 & 0.985 & 0.983 & 0.908 & 0.030 & 0.956 & 0.954 & 0.832 \\
\midrule
\textbf{DeRA-MOS (Full)} & 0.018 & \textbf{0.989} & \textbf{0.989} & \textbf{0.940} & \textbf{0.028} & \textbf{0.958} & \textbf{0.956} & \textbf{0.835} \\
\bottomrule
\end{tabular}
\vspace{-15pt}
\end{table*}

\section{Experimental Setups}

\subsection{Dataset and Evaluation Metrics}

Our DeRA-MOS framework was evaluated on the MusicEval dataset \cite{liu2025}, the official benchmark for the AudioMOS 2025 Challenge \cite{huang2025}.
The dataset comprised outputs from 31 distinct TTM systems generated from 384 fixed prompts, each annotated by experts for MI and TA.
Unlike previous work that relied on custom stratified splits for internal validation \cite{ritter-gutierrez2025}, we strictly adhered to the official train, development, and test data partitions to ensure standardized benchmarking.

% \subsection{Evaluation Metrics}

We evaluated performance using four metrics: SRCC, Kendall's Tau (KTAU) \cite{kendall1938}, linear correlation coefficient (LCC), and MSE \cite{cooper2022}.
While SRCC and KTAU were our primary metrics for validating the global ranking capability introduced by BALR, LCC and MSE were included to assess linear tracking and absolute prediction accuracy.
This evaluation-metric suite ensured that our decoupled framework not only ranked systems correctly but also maintained absolute score fidelity through SAMA's geometric anchoring.

\subsection{Implementation Details}

To rigorously isolate the effectiveness of our task-decoupled optimization strategy, we adopted the optimal dual-branch architecture from the DORA-MOS baseline \cite{ritter-gutierrez2025} as our fixed backbone.
Specifically, audio features were extracted using the frozen pre-trained MuQ model \cite{zhu2025}, and text embeddings were encoded via the frozen RoBERTa model \cite{liu2019}.
The temporal dependencies were modeled using a Transformer encoder paired with attention pooling, serving as a shared backbone before splitting into task-specific MI and TA prediction heads.
By fixing the architectural components, we minimized confounding factors so that observed gains could be attributed primarily to BALR and SAMA.
Applied immediately after the initial linear projections, $\mathcal{L}_{SAMA}$ imposes a pre-fusion geometric constraint.
It acts as a score-conditioned regularizer, pulling high-TA pairs closer while separating low-TA pairs, to preserve cross-modal structure and suppress representation drift.
Since $\mathcal{L}_{BALR}$ and $\mathcal{L}_{SAMA}$ serve purely as training regularizers, DeRA-MOS maintains the exact same inference complexity as the baseline (zero additional parameters and FLOPs), ensuring highly efficient deployment.

\subsection{Training Configuration}

The network was optimized using an AdamW optimizer \cite{loshchilov2019} with a learning rate of $5 \times 10^{-5}$.
Crucially, since the BALR loss conceptualizes the mini-batch as a query list for global sorting, we utilized a batch size of 32.
Under standard random shuffling, the selected batch size provided a sufficiently diverse distribution of MOS scores within each step to compute a meaningful ranking gradient.
For the hyperparameter settings, the temperature in the BALR softmax distribution was empirically set to $\tau = 1.0$.
The overall loss function integrated the task-decoupled objectives with weighting coefficients $\alpha = 0.2$ and $\beta = 0.3$, selected by development-set tuning.
The training process spanned a maximum of 100 epochs, with early stopping (patience: 15 epochs) based on validation loss to prevent overfitting.
Evaluating with five random seeds yielded highly stable MI and TA SRCCs ($0.989 \pm 0.002$ and $0.956 \pm 0.003$, respectively).
We also conducted significance testing on paired SRCC outcomes between DeRA-MOS and the reproduced baseline.

\section{Results and Discussion}

\subsection{Comparisons with State-of-the-Art Systems}

Table \ref{tab:main_results} presents a comprehensive comparison between our proposed framework and state-of-the-art systems from the AudioMOS 2025 Challenge \cite{huang2025}.
To ensure a rigorous and transparent evaluation, we included both the officially reported numbers (denoted with *) and our reproduced DORA-MOS baseline.
The minor performance drop of our reproduced baseline compared to the official report is attributed to our strict adherence to standardized data splits and unified hyperparameter tuning, necessary for a fair ablation analysis.
Although the MI SRCC improvement over the official DORA-MOS* is numerically small, DeRA-MOS matches this level of performance without increasing architectural complexity.
Crucially, compared with the reproduced baseline, it improves KTAU from $0.890$ to $0.940$ ($+0.050$), indicating substantially fewer rank inversions while preserving absolute score fidelity (MSE: $0.018$).
For TA, DeRA-MOS improves SRCC from $0.952$ to $0.956$ and reduces MSE from $0.060$ to $0.028$.
A system-level paired Wilcoxon signed-rank test \cite{woolson2008} shows that the MI/TA SRCC gains over the reproduced baseline are statistically significant ($p < 0.01$).

Beyond system-level evaluation, we also investigated the utterance-level performance to assess fine-grained quality ranking.
For music impression, DeRA-MOS achieves an utterance-level MI SRCC of $0.854$, outperforming the reproduced baseline ($0.845$).
This utterance-level MI gain indicates that BALR improves local relative quality ordering.
However, utterance-level TA SRCC decreases slightly ($0.632 \rightarrow 0.605$).
Although $\mathcal{L}_{SAMA}$ stabilizes global geometry, the batch-level pressure induced by $\mathcal{L}_{BALR}$ can oversmooth prompt-specific nuances.
Importantly, adjusting the $\alpha$ weight provides a controllable trade-off between global ordering and fine-grained local accuracy, which we plan to further optimize using multi-task gradient surgery \cite{yu2020}.

\begin{figure}[t]
\centering
\includegraphics[width=1.0\linewidth]{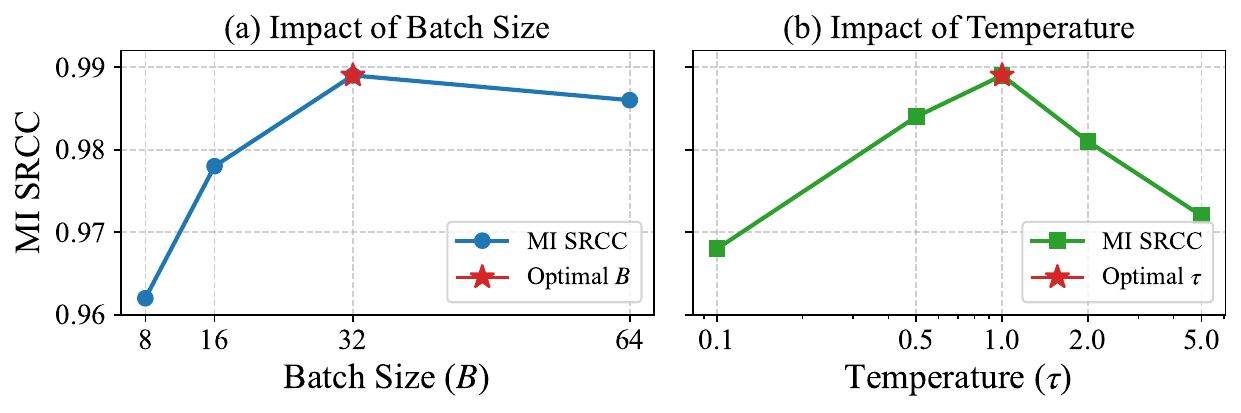}
\vspace{-25pt}
\caption{
Hyperparameter sensitivity analysis of DeRA-MOS on MusicEval.
(a) The impact of batch size ($B$) on MI SRCC.
(b) The impact of the temperature scaling factor ($\tau$) in the BALR loss on MI SRCC.
}
\label{fig:hyperparam}
\vspace{-15pt}
\end{figure}

\subsection{Ablation Study: The Synergy of Decoupling}

To unpack the contributions of our proposed objectives, we systematically ablated DeRA-MOS from our reproduced baseline (Table \ref{tab:main_results}, bottom section).
This ablation analysis isolates the role of each decoupled component.

\textbf{The cost of isolated ranking:}
Using only +Ranking ($\mathcal{L}_{BALR}$) improved MI ordering (KTAU: $0.890 \rightarrow 0.908$), confirming that listwise supervision was effective for rank consistency.
However, TA SRCC dropped ($0.952 \rightarrow 0.940$), indicating that ranking pressure alone could distort shared cross-modal representations.

\textbf{The role of geometric anchoring:}
Using only +Alignment ($\mathcal{L}_{SAMA}$) sharply improved TA calibration (MSE: $0.060 \rightarrow 0.030$) while preserving strong ranking metrics.
This ablation result confirmed that geometric anchoring enforced meaningful audio-text correspondence before fusion.

\textbf{Synergy in DeRA-MOS:}
Combining both objectives yielded the best overall profile.
The full model recovered MI calibration (MSE $0.018$), reached the highest MI KTAU ($0.940$), and improved TA robustness (MSE $0.028$, SRCC $0.956$).
These results showed that $\mathcal{L}_{BALR}$ and $\mathcal{L}_{SAMA}$ were complementary: BALR strengthened global ordering, while SAMA constrained geometry to prevent representation drift.

\subsection{Hyperparameter Sensitivity Analysis}

To demonstrate the robustness of our framework and validate the mathematical intuition behind the BALR loss, we analyzed the impact of its two core hyperparameters: batch size ($B$) and temperature ($\tau$).

\textbf{Impact of Batch Size:}
As shown in Fig. \ref{fig:hyperparam} (a), MI SRCC degrades noticeably when $B \le 16$, because a small mini-batch lacks sufficient distribution diversity to construct a meaningful query list, weakening the ranking gradient.
The performance peaks at $B=32$, providing adequate inter-sample dynamics, and saturates at $B=64$, confirming that a moderately large batch is essential for effective global ranking.

\textbf{Impact of Temperature Scaling:}
An excessively low temperature ($\tau=0.1$) sharpens the softmax distribution toward a one-hot vector, impeding gradient flow (Fig. \ref{fig:hyperparam} (b)).
Conversely, a high temperature ($\tau \ge 2.0$) overly smooths the probabilities, washing out subtle relative ranking signals among densely clustered MOS values.
The empirical optimum at $\tau=1.0$ strikes the best balance, providing theoretical and empirical stability by shaping a smooth yet discriminative probability distribution for the ranking landscape.

\textbf{Loss-Weighting Robustness:}
Beyond $B$ and $\tau$, we also examined the loss weights ($\alpha, \beta$).
Performance remained highly stable (MI SRCC variance $<0.002$) across settings with $\alpha \in [0.1, 0.3]$ and $\beta \in [0.2, 0.4]$, confirming that DeRA-MOS maintains robust decoupling without exhaustive weight tuning.

\begin{figure}[t]
\centering
\includegraphics[width=1.0\linewidth]{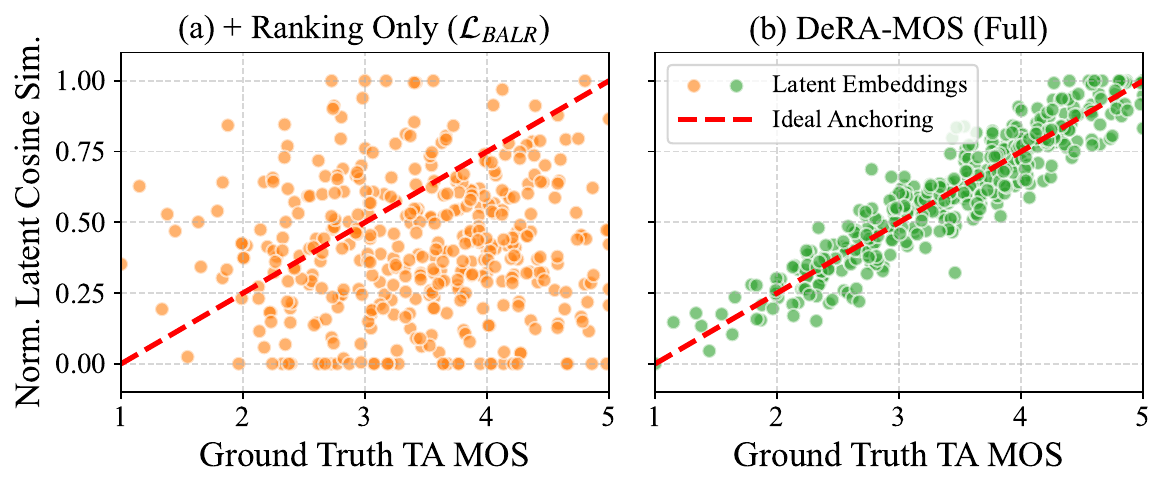}
\vspace{-25pt}
\caption{
Visualization of cross-modal representation drift and geometric anchoring.
(a) Ranking alone ($\mathcal{L}_{BALR}$) leads to unconstrained representation drift.
(b) Joint optimization with $\mathcal{L}_{SAMA}$ geometrically anchors the embeddings to the ideal trajectory (red dashed line).
}
\label{fig:drift}
\vspace{-15pt}
\end{figure}

\subsection{Latent Space Analysis: Visualizing Representation Drift}

To investigate the task interference observed in our ablation study, we visualized the latent cross-modal embedding space prior to cross-attention fusion.
Fig. \ref{fig:drift} plots the normalized latent cosine similarity $\tilde{c}^{(i)}$ (as described in Section \ref{sec:sama}) against the ground truth TA MOS $s_{TA}^{(i)}$.
The red dashed line represents the ideal anchoring trajectory defined in Eq. \ref{eq:sim}, which linearly maps the $[1, 5]$ subjective scores to the $[0, 1]$ geometric similarity space.
As clearly shown in Fig. \ref{fig:drift} (a), applying $\mathcal{L}_{BALR}$ without geometric constraints causes severe representation drift prior to cross-attention fusion.
The latent similarities scatter widely with virtually no correlation to the ground truth, confirming that isolated ranking optimization disrupts the shared cross-modal semantic space.
Conversely, Fig. \ref{fig:drift} (b) demonstrates the efficacy of the $\mathcal{L}_{SAMA}$ objective.
By explicitly penalizing deviations from the ideal trajectory, DeRA-MOS successfully anchors the previously drifting representations, forcing them to tightly converge along the red dashed line.
This drift visualization confirms that explicit geometric anchoring is essential to prevent modality collapse during listwise optimization.

\section{Conclusion and Future Work}

In this letter, we proposed DeRA-MOS, a task-decoupled framework for TTM MOS evaluation.
BALR aligns optimization with rank-based metrics, while SAMA enforces cross-modal geometric consistency.
On MusicEval, DeRA-MOS significantly outperformed the baseline in SRCC/KTAU and improved absolute TA accuracy.
Since both proposed losses are training-only, inference efficiency is preserved, making the framework ideal for large-scale model development.

Currently, our evaluation is limited to the MusicEval dataset due to a lack of other standardized TTM benchmarks. Future work will aim to validate this framework on broader, diverse datasets as they become available.
Furthermore, we plan to extend this paradigm to broader audio domains, including text-to-speech \cite{kim2021,kong2020} and zero-shot sound effects evaluation, and integrate probabilistic modeling to better account for annotator error rates and subjective uncertainty \cite{dawid1979}.

\newpage

\bibliographystyle{IEEEtran}
\bibliography{references.bib}
\end{document}